\documentclass[prb,aps,nobalancelastpage,amssymb,twocolumn,citeautoscript]{revtex4}

\usepackage{graphicx}

\begin{document}

\title{On the transport and thermodynamic properties of quasi-two-dimensional purple bronzes A$_{0.9}$Mo$_6$O$_{17}$ (A=Na, K)}

\author{Xiaofeng Xu$^{1,2,3}$, A. F. Bangura$^{2,4}$, C. Q. Niu$^1$, M.~Greenblatt$^5$, Song Yue$^6$, C. ~Panagopoulos$^3$ and N. E. Hussey$^2$}

\affiliation{$^1$Department of Physics, Hangzhou Normal University, Hangzhou 310036, China}
\affiliation{$^2$H. H. Wills Physics Laboratory, University of Bristol, Tyndall Avenue, BS8 1TL,
United Kingdom} \affiliation{$^3$Division of Physics and Applied Physics, School of Physical and
Mathematical Sciences, Nanyang Technological University, 637371 Singapore}
\affiliation{$^4$RIKEN(The Institute of Physical and Chemical Research) , Wako, Saitama 351-0198,
Japan.} \affiliation{$^5$Department of Chemistry and Chemical Biology, Rutgers University,
Piscataway, NJ 08854} \affiliation{$^6$Department of Physics, Jinan University, Guangzhou, China}

\date{\today}

\begin{abstract}
We report a comparative study of the specific heat, electrical resistivity and thermal conductivity
of the quasi-two-dimensional purple bronzes Na$_{0.9}$Mo$_6$O$_{17}$ and K$_{0.9}$Mo$_6$O$_{17}$,
with special emphasis on the behavior near their respective charge-density-wave transition
temperatures $T_P$. The contrasting behavior of both the transport and the thermodynamic properties
near $T_P$ is argued to arise predominantly from the different levels of intrinsic disorder
in the two systems. A significant proportion of the enhancement of the thermal conductivity above $T_P$ in
Na$_{0.9}$Mo$_6$O$_{17}$, and to a lesser extent in K$_{0.9}$Mo$_6$O$_{17}$, is attributed to the
emergence of phason excitations.
\end{abstract}

%\pacs{}%
\maketitle
\section{Introduction}
\label{Intro}

It is well known that in many low-dimensional compounds, Fermi surfaces are unstable against a
Peierls instability below a critical temperature, $T_P$. The mean-field description of this
so-called charge-density-wave (CDW) state is characterized by the opening of the electronic gap at
the Fermi level, accompanied by a softening in the phonon spectrum at $Q=2k_F$, with $k_F$ the
Fermi wave vector. As first pointed out by Lee, Rice and Anderson \cite{Lee74}, this soft-phonon
spectrum splits into two new modes of collective oscillations: one corresponds to optic-like
amplitude mode and the other one to an acoustic-like phase mode. This phase mode or phason existing
at zero frequency in the ideal CDW compounds corresponds to the sliding motion of the CDW as
proposed by Fr\"{o}hlich \cite{Frohlich}. In real materials, however, the phason excitations are
gapped out primarily due to impurity pinning. While these gapped phason excitations may not
contribute to the {\it charge} transport directly, they may still play a role in the {\it thermal} transport
owing to the finite value of $d\omega/dq$ in their spectrum.

The quasi-two-dimensional (Q2D) purple bronzes Na$_{0.9}$Mo$_6$O$_{17}$ and K$_{0.9}$Mo$_6$O$_{17}$
undergo CDW transitions at $T_P \sim$ 80K, 120K respectively, which has been confirmed by various
probes such as electrical resistivity and magnetic susceptibility\cite{Greenblatt88}. In
K$_{0.9}$Mo$_6$O$_{17}$, diffuse  X-ray scattering and electron diffraction experiments have also
revealed the three nesting wave vectors of ($a^\ast$, 0, 0) and its symmetrical equivalents below
$T_P$ \cite{Escribe-Filippini}, while STM studies further observed the (2$\times$2) superstructure
imposed on the crystal lattice at low temperatures \cite{Mallet}. With regard to their crystal
structures, these two compounds share a great deal of commonality albeit with slight differences in
symmetry, i.e., trigonal P\={3} symmetry in K$_{0.9}$Mo$_6$O$_{17}$ versus monoclinic C2 symmetry
in Na$_{0.9}$Mo$_6$O$_{17}$ \cite{Greenblatt88}. The main building block in both systems is the
slabs of Mo-O corner-sharing polyhedra, consisting of four MoO$_6$ octahedral layers terminated on
either side by a layer of MoO$_4$ tetrahedra. These slabs lie perpendicular to the $c$-axis and are
separated from each other by a layer of alkali metals. The MoO$_4$ tetrahedra in adjacent layers do
not share corners, disrupting the Mo-O-Mo bonding along the $c$-axis \cite{Greenblatt88}. This
layered structure is expected to lead to a Q2D electronic structure, as confirmed by dc transport
measurements \cite{Greenblatt88}.

The precise topology of the Na$_{0.9}$Mo$_6$O$_{17}$ Fermi surface (FS), as revealed by
angle-resolved photoemission spectroscopy (ARPES) \cite{Gweon, Breuer}, remains controversial, with
different ARPES groups claiming a FS topology that is either similar to \cite{Gweon} or
distinct from \cite{Breuer} that of K$_{0.9}$Mo$_6$O$_{17}$. According to Breuer {\it et al.}
\cite{Breuer}, there is only one electron pocket centered around the $\Gamma$-point in
Na$_{0.9}$Mo$_6$O$_{17}$, rather than the two pockets found in K$_{0.9}$Mo$_6$O$_{17}$. Moreover,
while the FS of K$_{0.9}$Mo$_6$O$_{17}$ can be viewed as a combination of  pairs of
quasi-one-dimensional (Q1D) Fermi sheets which can be nested to one another by the so-called
\lq hidden' nesting vectors \cite{Whangbo87, Whangbo91}, Breuer {\it et al.} find no evidence for the
1D FS parallel to the $\Gamma$X direction in sodium purple bronze. These subtle yet important
differences in the proposed electronic structure are supported, on a qualitative level at least, by
the observation that the thermoelectric power $S(T)$  below $T_P$ is dramatically different in the
two systems \cite{Tian97}; while $S(T)$ in both systems decreases linearly with decreasing
temperature towards zero at $T = T_P$, below $T_P$, $S(T)$ shows a large negative (positive) peak
for A = Na (K) respectively, suggesting that the dominant carrier has an opposite sign in the two
cases.

To gain further insight into the nature of the CDW formation in these two compounds, in particular
the changes in the nature of their electronic and phononic sub-systems at $T = T_P$, we present in
this paper a comparative study of the specific heat, electrical resistivity and thermal
conductivity of Na$_{0.9}$Mo$_6$O$_{17}$ and K$_{0.9}$Mo$_6$O$_{17}$ over a broad temperature range
above and below their respective CDW transitions. A number of key features are uncovered. Firstly,
the anisotropy of the electrical resistivity is found to be an order of magnitude higher than has
been reported in the existing literature. Secondly, in Na$_{0.9}$Mo$_6$O$_{17}$, the in-plane
thermal conductivity $\kappa_{ab}(T)$ is found to decrease across the CDW transition, in marked
contrast to the significant enhancement of $\kappa_{ab}(T)$ observed at $T_P$ in
K$_{0.9}$Mo$_6$O$_{17}$. Finally, in the heat capacity data, a large $\lambda$-shape anomaly is
observed in K$_{0.9}$Mo$_6$O$_{17}$, while in Na$_{0.9}$Mo$_6$O$_{17}$, the heat capacity exhibits
a broad hump with no discernible feature at $T_P$. We argue here that the principal origin of these
distinct physical responses in the two systems is the difference in their alkali-metal
stoichiometry. We also reveal the first possible manifestation of phason excitations in the thermal
conductivity of these archetypal CDW bronzes above $T_P$.

\section{Experimental}
\label{Exp}

The single crystals used in this study were grown by electrolytic reduction of a melt of
A$_2$CO$_3$-MoO$_3$ (A = Na,K) with an appropriate molar ratio. The details of the sample growth
procedure are described elsewhere \cite{Tian97, Tian01, Tian01b, Tian02, Greenblatt85}. Good single
crystallinity of the as-grown samples was confirmed by a single-crystal X-ray diffractometer. The
platelet-like samples were then cut into bar shapes of appropriate geometry for in-plane and
out-of-plane transport measurements. The resistivity of each sample was measured with a standard
four-probe $ac$ lock-in detection technique.

For the thermal conductivity measurements, we developed a zero-field apparatus housed in a $^4$He
flow cryostat that covers the temperature range 10K $<$ $T$ $<$ 300K. We employed a modified
steady-state method, shown schematically in Figure S2 of Ref. \cite{Wakeham11}, in which a
temperature gradient, measured using a differential thermocouple, is set up across the sample
through a pair of calibrated heat-links attached to each end. The sample is suspended by the free
ends of the heat-links between two platforms that are weakly coupled to the heat-bath. Each
platform houses a heater that enables a temperature gradient to be set up across the sample in both
directions at a fixed heat-bath temperature. The heat-links are also differential thermocouples
that allow the heat flowing into and out of the sample to be measured simultaneously in order to
determine the heat loss across the sample in steady state.

\begin{figure}
    \includegraphics[width=7cm,keepaspectratio=true]{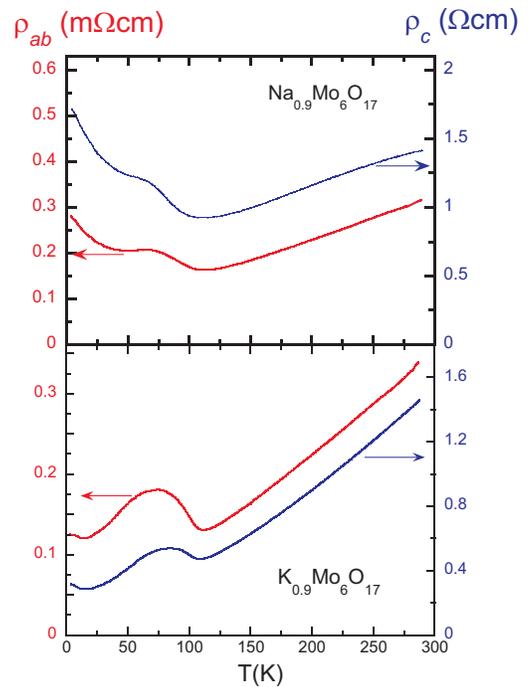}
        \caption{(Color online) The zero-field in-plane (left axis) and inter-plane (right axis)
resistivity of Na$_{0.9}$Mo$_6$O$_{17}$ (top panel) and K$_{0.9}$Mo$_6$O$_{17}$ (bottom panel) as a
function of temperature between 4.2K and 300K. The large anisotropy is evident from their different
$y$-axis scales.}
    \label{Fig1}
\end{figure}

The heat capacity measurements were performed in a commercial Quantum Design PPMS system. The
measurements were done on a large single-crystal piece of Na$_{0.9}$Mo$_6$O$_{17}$ (of 2.495mg in
weight), and two pieces of single-crystalline K$_{0.9}$Mo$_6$O$_{17}$ (total weight 1.190mg). The
heat capacity of the addenda was determined in a separate run and subtracted from the total heat
capacity. In the $C_p(H)$ measurements, a static magnetic field of 14 Tesla was applied along the
crystallographic $c$-axis.

\section{Results}
\label{Results}

\begin{figure}
    \includegraphics[width=9cm,keepaspectratio=true]{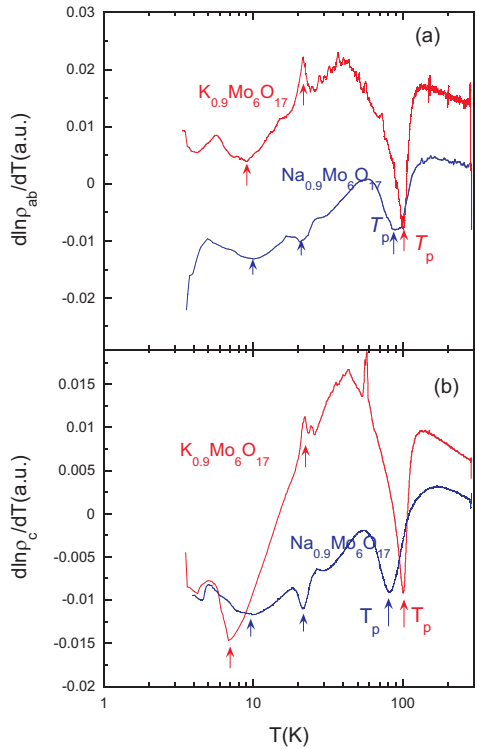}
        \caption{(Color online) Panels (a) and (b) plot d($\ln{\rho_{ab}})$/d$T$ and d($\ln{\rho_c})$/d$T$
respectively to identify the temperature and the width of the CDW transition. In addition to the
main CDW transition at $T_P$, there exist extra structures at low temperature, pointed out by the
color-coded arrows, which may be linked to other sub-dominant CDW/SDW transitions.}
    \label{Fig2}
\end{figure}

Fig.~\ref{Fig1} summarizes both the in-plane ($\rho_{ab}$) and inter-plane ($\rho_c$) resistivities
of our Na$_{0.9}$Mo$_6$O$_{17}$ and K$_{0.9}$Mo$_6$O$_{17}$ single crystals. Overall, the
temperature dependence in each case is in good agreement with what has been reported previously
\cite{Greenblatt88, Tian97}. Specifically, both $\rho_{ab}(T)$ and $\rho_c(T)$ are approximately
$T$-linear at high temperatures then pass through a well-defined minimum around $T = T_P$. At the
lowest temperature studied, both systems develop a second upturn that is larger in the  A = Na
bronze. In both cases, the zero-temperature resistivity tends towards a finite value, confirming
that these CDW transitions are metal-metal transitions. As the resistivity minima at $T_P$ in both
systems are rather broad, we use their derivatives d($\ln{\rho})$/d$T$, plotted in Fig.~\ref{Fig2},
to define the onset of the CDW transition, as is the norm \cite{Greenblatt88}. From both
d($\ln{\rho_{ab}})$/d$T$ and d($\ln{\rho_c})$/d$T$, deep minima are identified at $\sim$80K and
$\sim$100K for Na$_{0.9}$Mo$_6$O$_{17}$ and K$_{0.9}$Mo$_6$O$_{17}$ respectively, in good agreement
with previous studies \cite{Greenblatt88}. Evidently, the width of the derivative minimum at $T$ =
$T_P$ is much sharper in K$_{0.9}$Mo$_6$O$_{17}$ than in Na$_{0.9}$Mo$_6$O$_{17}$, suggestive of a
lower intrinsic disorder level in the former. We shall return to this point in the Discussion
section.

In addition to the similarities highlighted above, there are also some notable differences between
our measurements and those reported in the literature. While the $\rho_c$ values of both systems
are comparable with earlier reports, the absolute values of $\rho_{ab}$ are roughly one order of
magnitude lower, making the corresponding resistivity anisotropy ($\sim$ 3000 - 4000) significantly
higher than previously reported. We attribute these lower in-plane resistivity values to an
improved shorting out of the highly resistive inter-plane current paths. This was achieved by
ensuring that the voltage and current contact pads cover the entire thickness of the crystal. If
this is not done carefully, any current flow perpendicular to the conducting planes will lead to a
much larger apparent resistivity being measured. We have used this method previously to
successfully isolate the low resistivity direction in a variety of low dimensional materials and
refer the interested reader to our recent work on the Q1D superconductor from the same family
(Li$_{0.9}$Mo$_6$O$_{17}$), in which we again found a significantly lower resistivity than had been
previously reported \cite{Mercure11}. Significantly, in that case, the resulting in-plane resistive
anisotropy (i.e. for currents parallel and perpendicular to the conducting chains) was found to be
consistent with that obtained independently from both optical spectroscopy measurements and the
measured upper critical field anisotropy. Such good agreement highlights the efficacy of our
technique in isolating the individual components of the conductivity tensor of highly anisotropic
systems.

\begin{figure}
\includegraphics[width=9cm,keepaspectratio=true]{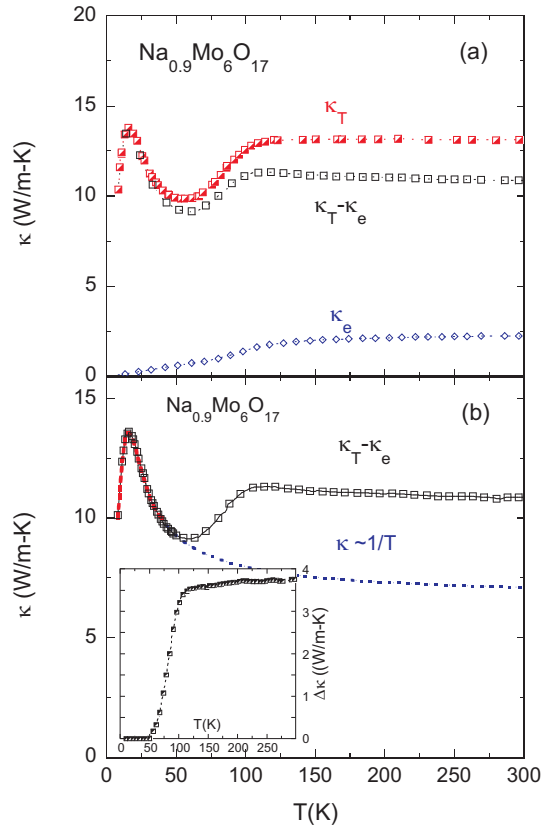}
\caption{(Color online) The red two-tone squares in (a) represent the temperature dependence of the
total thermal conductivity of Na$_{0.9}$Mo$_6$O$_{17}$, as measured directly from the experiments.
The blue open circles give the electronic part as determined from their electrical conductivity and
the Wiedemann-Franz law, leaving the residual part, $\kappa_T-\kappa_e$, as indicated by the black
open squares and replotted in panel (b) accordingly, correspond to the contributions from all other
excitations. The red solid line in panel (b) indicates the fits to the phonon Boltzmann transport
theory (see Text). The phonon thermal conductivity $\kappa_{ph}$, indicated by the blue dotted
line, evolves as $\sim 1/T$ at high temperature. A well defined excess to $\kappa_{ph}$, as plotted
in the inset, is clearly evident around and above $T_P$.} \label{Fig3}
\end{figure}

The origin of the second resistive upturn in both bronzes at low temperatures is still unresolved,
with spin density wave (SDW) formation, a second CDW transition and localization all put forward as
possibilities. Significantly, in Li$_{0.9}$Mo$_6$O$_{17}$, a similar upturn is also
seen \cite{Greenblatt84}, though in this case, the excellent scaling observed in the longitudinal
intrachain magnetoresistance \cite{Xu09} is considered to point more towards some form of DW
gapping than localization.

\begin{figure}
\includegraphics[width=9cm,keepaspectratio=true]{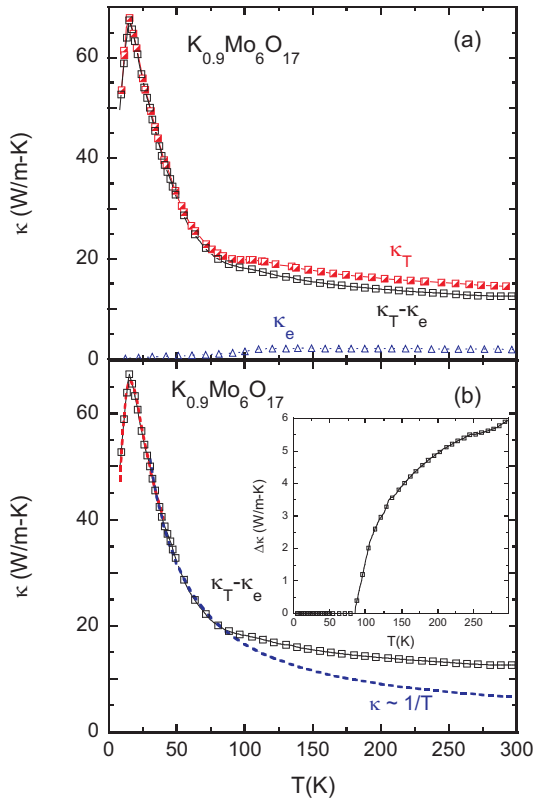}
\caption{(Color online) Analogous to Fig. \ref{Fig3}, the total thermal conductivity $\kappa_T$ of
K$_{0.9}$Mo$_6$O$_{17}$ as measured from the experiments is decomposed into the electronic part
$\kappa_e$ and the residual part $\kappa_T-\kappa_e$ in panel (a). Again, the phonon Boltzmann
transport theory was applied to fit the low-$T$ $\kappa_T-\kappa_e$ data, as indicated by the red
solid curve in panel (b). A well defined excess to the phonon thermal conductivity $\kappa_{ph}$,
which is characteristic of $\sim 1/T$ decay at high temperature and shown by the blue dashed line,
is plotted in the inset and evidently seen around and above $T_P$.} \label{Fig4}
\end{figure}

The total in-plane thermal conductivity $\kappa_{ab}$ as a function of temperature is plotted as
red two-tone squares in the panels (a) of Fig.~\ref{Fig3} and Fig.~\ref{Fig4} for
Na$_{0.9}$Mo$_6$O$_{17}$ and K$_{0.9}$Mo$_6$O$_{17}$ respectively. Compared to the zero-field
resistivities, which have similar magnitudes and overall display similar behavior (bar the size of
the resistive upturn), the in-plane thermal conductivities of the two systems show some notable
differences. In Na$_{0.9}$Mo$_6$O$_{17}$, $\kappa_{ab}(T)$ is nearly $T$-independent above $T_P$,
then as $T$ cools below $T_P$, $\kappa_{ab}(T)$ first develops a well-defined downturn, goes
through a minimum around 50K then finally peaks at around 15 K. The thermal conductivity of
K$_{0.9}$Mo$_6$O$_{17}$, on the other hand, while showing a slight kink at $T$ $\sim$ 120 K,
increases significantly below $T_P$, reaching a peak value at $T$ = 20K that is approximately four
times higher than that of Na$_{0.9}$Mo$_6$O$_{17}$. The low temperature peak is typical of the
thermal response of phonons and results from the competition between increasing numbers of
propagating phonons (with increasing $T$) and the concomitant reduction of their mean free path.
The larger magnitude of the peak amplitude of $\kappa_{ab}$ below $ T_P$ in K$_{0.9}$Mo$_6$O$_{17}$
compared with Na$_{0.9}$Mo$_6$O$_{17}$ again reflects a lower level of static disorder in the
former.

The total thermal conductivity of a metal consists of the contributions from free carriers, lattice
vibrations and other possible excitations. The electronic component $\kappa_e$ in an ordinary metal
can be estimated from the Wiedemann-Franz law \cite{Ashcroft}

\begin{eqnarray}
\kappa_e/\sigma T = L_0 \label{eqn:one}
\end{eqnarray}

\noindent where $\sigma$ is the dc electrical conductivity and $L_0$=2.45$\times$10$^{-8}$
W$\Omega$K$^{-2}$ is the so-called Lorenz number. Although deviations in this relation are observed
in most metals at intermediate temperatures, due to the differing effects of small- and large-angle
scattering on the electrical and thermal conductivities, these are typically only of order 20-30\%
and as a first approximation, they can be ignored.

The blue open circles in panel (a) of Fig.~\ref{Fig3} and Fig.~\ref{Fig4} represent the electronic
contribution $\kappa_e$, calculated from Eqn.~(1) and the electrical resistivity data plotted in
Fig.~\ref{Fig1}. The kinks in $\kappa_e$ of both systems at $T = T_P$ result primarily from the
substantial decrease of the carrier density at the CDW transition. The residual part,
$\kappa_T-\kappa_e$, as replotted in (b) of Fig.~\ref{Fig3} and Fig.~\ref{Fig4} accordingly, thus
represents the contributions to the thermal conductivity arising from phonons and any other
excitations. Note that the slight kink in $\kappa_{ab}(T)$ of K$_{0.9}$Mo$_6$O$_{17}$ is
effectively removed once $\kappa_e$ is subtracted from the raw  data. By contrast, for
Na$_{0.9}$Mo$_6$O$_{17}$, the kink is still very much evident even after subtraction of $\kappa_e$.
It is important to note that this difference is not simply due to our overestimate of
$\kappa_{ab}$, since we obtained $\kappa_{ab}$ values of a similar magnitude on a second crystal,
nor to any possible underestimate of $\sigma_{ab}$. Indeed, in order to remove the kink in
$\kappa_T-\kappa_e$ completely, we would have had to underestimate $\sigma_{ab}$ by a factor of 5,
i.e. by an amount more than one order of magnitude greater than our experimental uncertainty (of
order 30\%). Moreover, the magnitude of $\sigma_{ab}$ that we measure is already one order of
magnitude higher than in previous reports. We are confident therefore that this kink is an
intrinsic feature of $\kappa_{ab}(T)$ in Na$_{0.9}$Mo$_6$O$_{17}$.

The main component of $\kappa_T-\kappa_e$ is the contribution from the phonons. In our subsequent
analysis, we employ the Debye model for the phonon Boltzmann transport theory \cite{Berman}:

\begin{eqnarray}
\kappa_{ph} =\frac{1}{2\pi^2 \upsilon} \int ^{\omega_{max}}_0 \hbar \omega^3 \tau
\frac{(\hbar\omega /k_B T^2)exp(\hbar\omega /k_B T)} {[exp(\hbar\omega /k_B T)-1]^2 } d\omega.
\label{EQN:two}
\end{eqnarray}

\noindent where $\omega$ is the phonon frequency, $\omega_{max}$ is the maximum phonon frequency
related to the Debye temperature $\Theta_D$ via $\hbar \omega_{max}=k_B\Theta_D$, $\upsilon =
\Theta_D(k_B/\hbar)(6\pi^2n)^{-1/3}$ is the mean sound velocity with $n$ the number density of
atoms, while $\tau$ is the phonon relaxation time which takes account of all the phonon scattering
mechanisms. We assume here that these individual scattering processes act independently so that
$\tau_\Sigma^{-1}=\Sigma \tau_i^{-1}$, where $\tau_i$ corresponds to each individual relaxation
time. The main phonon scattering mechanisms come from scattering off two-dimensional defects and
phonon-phonon scattering, that is \cite{Kordonis}

\begin{eqnarray}
\tau_\Sigma^{-1}=A\omega^2+B\omega^2T exp(-\Theta_D/bT). \label{EQN:three}
\end{eqnarray}

\noindent In the same way that the quasiparticle mean-free-path $\ell$ cannot become shorter than
the interatomic spacing $a$ (the so-called Mott-Ioffe-Regel limit \cite{Hussey04}), we follow the
procedure of Kordonis {\it et al.} and assume here that the {\it phonon} mean-free-path must also
have a lower limit $\ell_{\texttt{min}}$ with
$\tau$=max($\tau_\Sigma$,$\ell_{\texttt{min}}/\upsilon$) \cite{Kordonis}. We thus set the
parameters $A$, $B$ and $b$ and $\ell_{\texttt{min}}$  as freely adjustable parameters and use the
Debye temperature extracted from low-$T$ heat capacity data, as shown below, to  obtain the
resultant fitting curves displayed as solid red lines in Fig.~\ref{Fig3}(b) and Fig.~\ref{Fig4}(b)
for Na$_{0.9}$Mo$_6$O$_{17}$ and K$_{0.9}$Mo$_6$O$_{17}$ respectively. The corresponding fitting
parameters are listed in Table I.  The values for $A$, $B$ and $b$ are typical for $\kappa_{ph}$ in
oxide systems. The values for $\ell{\texttt{min}}$, on the other hand, are rather large, though as
mentioned by Kordonis {\it et al.} \cite{Kordonis}, the inclusion of $\ell{\texttt{min}}$ has
little bearing on the fitting of the low-temperature $\kappa_{ph}(T)$. Above the fitting range, we
assume the typical $1/T$ decay in the phonon thermal conductivity due to phonon-phonon scattering.
These tails are depicted in Fig.~\ref{Fig3}(b) and Fig.~\ref{Fig4}(b) as dashed blue lines.

\begin{table}
\begin{center}
\caption{Fitting parameters to Eqn. \ref{EQN:two} with the relaxation time of the form of Eqn.
\ref{EQN:three}. $\ell_{\texttt{min}}$ corresponds to the low boundary of phonon mean free path
(see Text).}
\begin{tabular}{l | c | c }
\hline\hline
Parameter & Na$_{0.9}$Mo$_6$O$_{17}$ & K$_{0.9}$Mo$_6$O$_{17}$\\
\hline
\textit{A}($10^{-16}$s) & 0.68 & 0.25\\
\textit{B} ($10^{-17}$ sK$^{-1}$)& 1.69 & 0.76\\
\textit{b} & 10.29 & 6.06 \\
$\ell_{\texttt{min}}$ ($\AA$) & 158 & 208 \\
Temperature region (K) & 8-40 & 8-50\\
\hline
\end{tabular}\end{center}

\label{TAB:fittingBoltzmann}
\end{table}

As mentioned above, the most striking feature of the $\kappa_T - \kappa_e$ data for
Na$_{0.9}$Mo$_6$O$_{17}$ shown in Fig.~\ref{Fig3}(b) is the enhancement of the thermal conductivity
beginning at around $T$ = 50 K. This enhanced conductivity, plotted in the inset of panel (b), is
reminiscent of that seen in other CDW systems such as (NbSe$_4$)$_{10}$I$_3$, K$_{0.3}$MoO$_3$ and
(TaSe$_4$)$_2$I \cite{Kwok89, Smontara93}, for which it was attributed to the emergence of phason
excitations associated with the CDW transition. Note that in these other CDW systems, this
contribution exists both above and below $T_P$ where it typically peaks. The contribution above
$T_P$ is considered to arise from fluctuations associated with the phason excitations. Moreover, as
noted by Smontara {\it et al.} \cite{Smontara93}, the phason fluctuations may extend to very high
temperature due to slow decay of the damping rate $\Gamma$. Hence, while we do not expect that
phason excitations account for the full excess of thermal conductivity in Na$_{0.9}$Mo$_6$O$_{17}$,
they may well account for a significant proportion of it. A similar enhancement can also be
inferred, though less convincingly, in the thermal conductivity data of K$_{0.9}$Mo$_6$O$_{17}$
plotted in the inset of Fig.~\ref{Fig4}(b) from extrapolation of the 1/$T$ decay to higher
temperature (dotted curve in Fig.~\ref{Fig4}(b)). Note that it is not possible to fit the full
$\kappa(T)$ data of K$_{0.9}$Mo$_6$O$_{17}$ using Eqn.~(2) and that a fit to the high-$T$
$\kappa(T)$ data incorporating $\ell_{\texttt{min}}$ gives unphysical values for $A$, $B$ and $b$
and strongly underestimates the enhancement in $\kappa(T)$ below $T_P$.

\begin{figure}
\includegraphics[width=9cm,keepaspectratio=true]{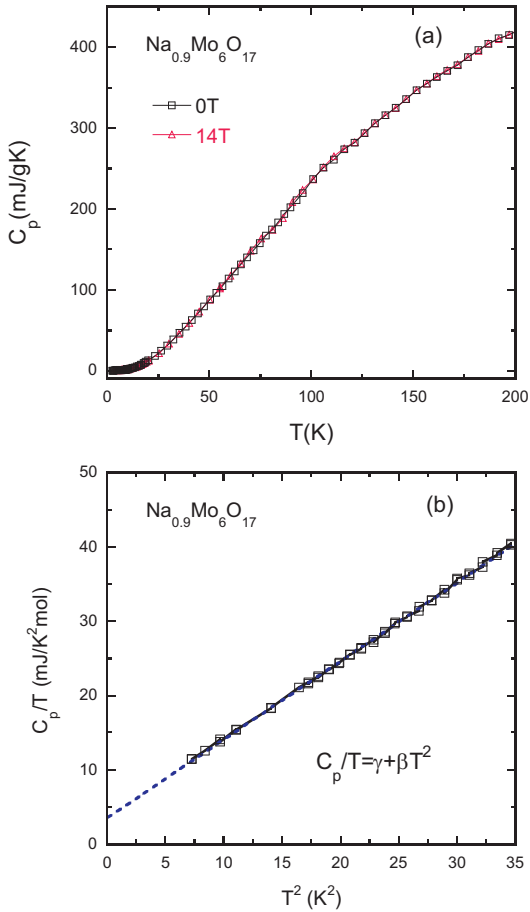}
\caption{(Color online) Panel (a): The temperature dependence of the specific heat in both zero
field (black square) and 14 Tesla magnetic field (red triangle) of Na$_{0.9}$Mo$_6$O$_{17}$.
Evidently, an applied magnetic field has little effect on its overall heat capacity. Panel (b): The
low- temperature zero-field $C_P/T$ data plotted as a linear function of $T^2$ to allow the normal
electronic and acoustic phonon contributions to the specific heat to be delineated. The fitting
gives $\gamma$=3.57 mJ/K$^2$mol, $\beta$=1.05 mJ/K$^4$mol.} \label{Fig5}
\end{figure}

\begin{figure}
\includegraphics[width=9cm,keepaspectratio=true]{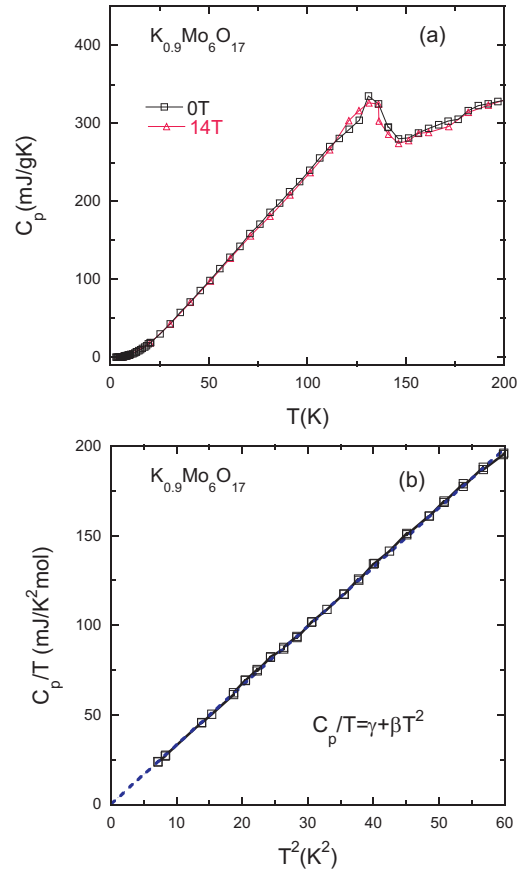}
\caption{(Color online) Similar to Fig. \ref{Fig5}, panel (a) shows the temperature dependence of
the specific heat in both zero field (black square) and 14 Tesla magnetic field (red triangle) of
K$_{0.9}$Mo$_6$O$_{17}$. As noted, an applied magnetic field has little effect on the total heat
capacity. Panel (b) separates the normal electronic and acoustic phonon contributions by fitting
$C_P/T$ of $H$=0T data as a linear function of $T^2$ at very low temperatures. The resultant
fitting parameters are $\gamma$=0.65 mJ/K$^2$mol, $\beta$=3.30 mJ/K$^4$mol.} \label{Fig6}
\end{figure}

The specific heat $C_p(T)$ data, shown in Fig.~\ref{Fig5} and Fig.~\ref{Fig6}, uncover another
marked difference between the two systems. For K$_{0.9}$Mo$_6$O$_{17}$, a large, well-defined heat
capacity anomaly, clearly associated with the CDW transition, is observed. The anomaly shown in
Fig.~\ref{Fig6}(a) is in fact much sharper than has been reported previously
\cite{Escribe-Filippini}. By contrast, the corresponding feature in Na$_{0.9}$Mo$_6$O$_{17}$ is
extremely broad and as such, is difficult to pinpoint. In contrast with the resistivity data, no
additional anomalies could be found at lower temperatures in either system due to any putative
sub-dominant CDW/SDW transitions. A magnetic field of 14 Tesla is found to have no effect on the
shape or position of the specific heat anomaly in both systems.

Plots of $C_p/T$ versus $T^2$ at the lowest temperature allow us to separate out the electronic ($=
\gamma T$) and phononic (= $\beta T^3$) contributions, as shown in panels (b) of Fig.~\ref{Fig5}
and Fig.~\ref{Fig6}. From the extracted $\gamma$ values, given in the Figure caption, we determine
the density of states of normal carriers at the Fermi level to be 1.515 eV$^{-1}$ per molecule and
0.275 eV$^{-1}$ per molecule for Na$_{0.9}$Mo$_6$O$_{17}$ and K$_{0.9}$Mo$_6$O$_{17}$,
respectively. Debye temperatures of 333K and 228K for Na$_{0.9}$Mo$_6$O$_{17}$ and
K$_{0.9}$Mo$_6$O$_{17}$ respectively can also be estimated from the formula $\Theta_D^3 =
\frac{5}{12}\pi^4rR\beta^{-1}$ with $R$ the gas constant and $r$ the total number of atoms in one
formula unit ($r$=24 for both cases). We note that the $\Theta_D$ value extracted here for
K$_{0.9}$Mo$_6$O$_{17}$ is somewhat lower than those reported in the literature \cite{Wang06,
Buder}.

\section{Discussions and Conclusion}
\label{Discussions}

In many CDW compounds, the thermal conductivity is enhanced substantially below the Peierls
transition, as a result of changes in the electron- or phonon-scattering processes. However, in
other CDW compounds, such as (NbSe$_4$)$_{10}$I$_3$, K$_{0.3}$MoO$_3$ and (TaSe$_4$)$_2$I
\cite{Kwok89, Smontara93}, the thermal conductivity is observed to pass through a well-defined
minimum below the Peierls transition before increasing again at lower temperatures. It has been
proposed that this additional contribution results from phason modes with large velocity
\cite{Smontara93}. Given the strong similarity in the behavior of $\kappa_{ab}(T)$ in
Na$_{0.9}$Mo$_6$O$_{17}$ with those systems listed above, it would seem appropriate to ascribe at
least part of the excess thermal conductivity shown in the inset of Fig.~\ref{Fig3}(b) to the same
origin.

While there is no well-defined minimum in $\kappa_{ab}(T)$ in K$_{0.9}$Mo$_6$O$_{17}$ in the region
of the Peierls transition, inspection of Fig.~\ref{Fig4}(b) reveals a small excess thermal
conductivity above the high-$T$ $1/T$ decay of $\kappa_{ph}$ that could correspond to the same
phason contribution seen much more clearly in Na$_{0.9}$Mo$_6$O$_{17}$. Evidence for phason
excitations from measurements of the low-temperature specific heat of K$_{0.9}$Mo$_6$O$_{17}$ was
claimed in Ref.~\cite{Wang06}, through the observation of a broad hump in $C/T^3$ around $T$ = 16
K. Given that this temperature coincides with the onset of the low-temperature upturn in the
resistivity of K$_{0.9}$Mo$_6$O$_{17}$, it is not yet clear whether this feature can be ascribed
unambiguously to phasons or to a second density-wave transition.

The origin of the striking difference in the thermoelectric power of the Q2D purple bronzes below
$T$ = $T_P$, as described briefly in the Introduction, remains to be resolved, though it is likely
to arise from subtle differences in the electronic structure, as hinted at by ARPES \cite{Gweon,
Breuer}, as well as in the relative mobilities of the carriers on the remnant Fermi surface(s).
According to Breuer {\it et al.} \cite{Breuer}, the Fermi surface in Na$_{0.9}$Mo$_6$O$_{17}$
comprises an elliptical electron pocket, i.e. around the $\Gamma$ point, that is less prone to
nesting, and a diamond-shaped hole pocket with flat regions that are highly susceptible to nesting.
The negative thermopower below $T_P$ in Na$_{0.9}$Mo$_6$O$_{17}$ is therefore consistent with the
notion that the remnant Fermi surface below $T_P$ will be predominantly located about the
more-rounded sections of the single electron pocket.

The second electron pocket found in K$_{0.9}$Mo$_6$O$_{17}$ uncovered by Gweon {\it et al.}
\cite{Gweon}, is star-like. Intriguingly, between the apices of the stars, lie rounded regions of
Fermi surface with {\it negative} curvature, i.e. that are centered around the edges of the
Brillouin zone rather than around its center. Hence, while the second pocket in
K$_{0.9}$Mo$_6$O$_{17}$ is electron-like, those regions  unaffected by the CDW transition will have
predominantly hole-like character. In the heavily overdoped cuprate La$_{2-x}$Sr$_x$CuO$_4$ ($x$ =
0.30), which also has an electron pocket with regions of negative curvature, a similar inversion of
the sign of the thermopower\cite{Nakamura93} and of the Hall coefficient \cite{Narduzzo08} is
observed.  In this regard, it would be very interesting to explore the evolution of the in-plane
Hall coefficient in both purple bronzes above and below $T_P$.

One of the most notable findings of this study is the observation that the Q2D purple bronzes
Na$_{0.9}$Mo$_6$O$_{17}$ and K$_{0.9}$Mo$_6$O$_{17}$ display distinct transport and thermodynamic
properties associated with their CDW transitions. Given the possible differences in the
Fermiologies of the two systems, as inferred by ARPES \cite{Gweon, Breuer}, and their very
different thermoelectric responses below $T$ = $T_P$ \cite{Tian97}, as described above, these
differences might, on first inspection, suggest that the nature and mechanisms of CDW formation in
these two Q2D bronzes are in fact distinct. However, we argue here that a more consistent
explanation of these contrasting behaviors can be provided by considering the different crystal
chemistries of the two systems and the resulting variation in disorder content.

Regarding the latter, we note from the high-temperature $T$-linear resistivities plotted in Fig.
\ref{Fig1} that the extrapolated residual resistivity in Na$_{0.9}$Mo$_6$O$_{17}$ is significantly
higher than in K$_{0.9}$Mo$_6$O$_{17}$. Moreover, the low-$T$ resistivity in the Na-bronze is
approximately twice that at the resistive minimum, while in K$_{0.9}$Mo$_6$O$_{17}$, it remains
comparable, even though the low-$T$ heat capacity data reveal that the density of states in
Na$_{0.9}$Mo$_6$O$_{17}$ is substantially higher below $T_P$. This comparison implies that the A =
Na bronze is significantly more susceptible to localization effects, an indicator of higher levels
of disorder scattering. This conclusion is further corroborated by the observation that the peak in
the phonon thermal conductivity at low temperatures, a quantitative measure of the phonon
mean-free-path, is approximately four times lower in Na$_{0.9}$Mo$_6$O$_{17}$ than in
K$_{0.9}$Mo$_6$O$_{17}$. This argument is also consistent with what was seen in Fig. \ref{Fig2}
where the width of negative peak of d($\ln{\rho})$/d$T$ at $T_P$, a strong indicator of disorder
level, is much narrower in K$_{0.9}$Mo$_6$O$_{17}$. Finally, the anomaly in the heat capacity,
while sharp in our K$_{0.9}$Mo$_6$O$_{17}$ crystal, is smeared out in the Na-doped counterpart. A
similarly broad heat capacity anomaly was also reported for K$_{0.9}$Mo$_6$O$_{17}$ almost three
decades ago \cite{Escribe-Filippini}, in crystals grown by a different group, though unfortunately,
there is no accompanying transport data in that study to allow us to compare their residual
resistivities directly. Nevertheless, such variation in the heat capacity anomaly in what are
nominally the same compound does suggest that there is indeed a strong sample dependence in the
thermodynamic properties of the two systems.

The Na$_{0.9}$Mo$_6$O$_{17}$ system is known to possess a variable Na content
\cite{Ramanujachary84}. Variation in stoichiometry in the sodium purple bronzes is usually
considered to be a characteristic that exists between crystals, rather than within an individual
crystal. However, given the higher mobility of sodium atoms within the melt relative to that of the
potassium, we speculate here that there may be more of a tendency for the Na ions to form clusters
within each crystal, which in turn will lead to enhanced elastic scattering of the quasiparticles
on the conducting molybdate chains. It is also noted that clustering of Na ions may lead to a
distribution of transition temperatures and therefore increase the breadth of the transition, as
indeed seen in Fig. \ref{Fig2}. However, a distribution of concentration could yield a distribution
of the Fermi vectors in different parts of the crystal, favoring the incommensurate character of
the CDW, a necessary condition to have phasons which does not seem to be fulfilled in
K$_{0.9}$Mo$_6$O$_{17}$.

In summary, we have studied the transport and thermodynamic properties of two Q2D purple bronzes
Na$_{0.9}$Mo$_6$O$_{17}$ and K$_{0.9}$Mo$_6$O$_{17}$ and have revealed significant differences in
their physical properties above and below their CDW transitions. These differences appear to arise
from differences in their stoichiometry, possibly caused by the volatility of sodium during crystal
growth. An unusual enhancement of the thermal conductivity has been attributed, in comparison with
other CDW systems, to the possible emergence of phason excitations in these two systems. Finally,
measurements of the electrical resistivity reveal an electrical anisotropy that is one order of
magnitude larger than previously thought.

\begin{acknowledgments}
The authors would like to acknowledge technical assistance from B. Fauqu\'{e}, X. Y. Tee and
helpful discussions with J. D. Fletcher and N. Shannon. This work was supported by the National
Natural Science Foundation of China(No. 11104051), the EPSRC (UK), the Royal Society and the
National Research Foundation, Singapore.
\end{acknowledgments}

\bibliography{NaKMoO}

%%%%%%%%%%%%%%%%%%%%%%%%%%%%%%%%%%%%%%%%%%%%%%%%%%%%%%

\end{document}